\documentstyle[11pt,paspconf,epsf]{article}

\begin{document}

\title{What's Emitting the Broad Emission Lines?}
\author{Kirk Korista}
\affil{Western Michigan University, Department of Physics, Kalamazoo,
MI 49008}

\begin{abstract}
I present a brief review about our ideas concerning the origin and
nature of the broad line emitting gas. This is one of the outstanding
problems in quasar research. I suggest the establishment of a central
data base of high quality quasar spectra as a means to realize an
understanding of quasars.
\end{abstract}


\section{Introduction}

{\em Question:} What is the broad line region (BLR), that is, what
would we see if we could image\footnotemark\/ it? {\em Answer:} Nobody
knows. Since the identification of their emission lines by Schmidt in
1963 we have wondered about this question. It is disturbing that the
origin of the line emitting gas remains to this day a large question
mark. There is no single working model. Our only framework from which
to begin our understanding is this: gas of unknown origin, with
particle densities comparable to those in the solar chromosphere, lies
distributed and moves within the gravitational potential of a central
supermassive black hole (Blandford \& Rees 1978; Rees 1984), and is
photoionized by the central ionizing continuum source (Davidson \&
Netzer 1979). The gas elemental abundances in even high redshift
quasars is at least solar (Hamann \& Ferland 1993; Hamann \& Ferland
1999; Hamann, these proceedings). But what the emitting gas is doing
and how it got there remain only loosely constrained.

I review briefly the various models, ideas ($\S~2$), and few
constraints ($\S~3$) concerning the origin and nature of the broad line
emitting gas (BLEG). In $\S~4$ I sum up the current situation, and
suggest the importance of the establishment of a central data base of
high quality quasar spectra as a means to realize an understanding of
quasars --- drawing parallels to our understanding of stars and their
spectra beginning one century ago.

\footnotetext{The angular diameter of the best studied BLR, that
belonging to NGC~5548, is $\sim$~0.2 milliarcseconds.}

\section{The Ideas}

Most of our ideas concerning the nature of the BLEG center around three
possibilities: clouds, accretion disks, and stars. Some models invoke
combinations or hybrids of these three. I will discuss each of these in
turn.

\subsection{Clouds}

Here I define a cloud to be a localized, non-self-gravitating entity
such as that found in the interstellar medium (ISM), existing
independently from accretion disks or stars. Debris from disks or stars
are discussed in their own sections below. As will become apparent, the
cloud models suffer from a variety problems, especially those involving
pressure equilibrium (confinement) and dynamical stability, and thus
may be transient ($\tau_{life} \leq \tau_{dyn}$) entities if they can
exist at all.

One of the more popular cloud models (McCray 1979; Krolik, McKee, \&
Tarter 1981) had its origins with the old two-phase equilibrium
scenario of the local Galactic ISM (Spitzer 1978). The clouds are
``cold'', dense ($\sim$~$10^{10}$~cm$^{-3}$) condensates in pressure
equilibrium within a hot ($\rm{T \approx 10^8}$~K), low density flow.
This hot phase was proposed to be in Compton equilibrium with the
quasar radiation field.  However, Mathews \& Ferland (1987; MF87)
punched several holes into this picture. First, it was determined that
the quasar radiation field's Compton temperature is only $\approx
10^7$~K (see also Fabian et al.\ 1986), far too low for the clouds to
be stable, or for the hot phase to be optically thin to X-ray
radiation. Second, even if the hot phase temperature were $10^8$~K drag
forces would impose a terminal velocity on the clouds far too small to
account for the emission line widths.  These problems could be overcome
if the temperature of the intercloud medium exceeded $10^9$~K, a
relativistic wind. However, MF87 also suggested that the cold-phase
cloud condensation process could be disrupted by various strong radial
forces present in the quasar environment. Internal line radiation
pressure can also be disruptive of clouds (Elitzur \& Ferland 1986). A
good review of the ``cloud'' stability problem is given in Mathews \&
Capriotti (1985).

Radiatively accelerated pancake clouds and clouds oscillating about an
equilibrium radius (gravitational vs.\ radiative accelerations) in an
unspecified confining quasar atmosphere have been proposed by Mathews
(1982; 1993). To obviate the requirement of a Compton heated medium,
Martin Rees (1987) proposed that the clouds could be magnetically
confined ($B \sim 1$~Gauss), similar to the situation in the filaments
of the Crab Nebula.

Confinement and stability problems aside, a number of investigators
have attempted to set limits on the minimum numbers of clouds required
to produce the observed smooth emission line profiles. The simulations
of Capriotti, Foltz, \& Byard (1981b) coupled with medium S/N ratio and
spectral resolution observations of Atwood, Baldwin, \& Carswell (1982)
placed the lower limit at $N_{cl} > 5 \times 10^4$. This estimate
assumed a simple local line profile with a thermal width corresponding
to a temperature $T \approx 10^4$~K. Arav et al.\ (1998a,b) used the
power of the Keck~I 10m telescope and a spectral resolution of 6~km/s,
with a similar cross-correlation analysis but more sophisticated
simulations of possible local line profiles, to push the lower limits
to $N_{cl} > 3 \times 10^7$, given certain assumptions. A question that
arises immediately is what process could give rise to so many clouds?
The lower limit could be as small as $N_{cl} > 3 \times 10^5$ if the
local line widths are significantly larger than 80 km/s. The lower
limit to $N_{cl}$ diminishes with increasing local line width. This
question of cloud number also applies to the scenario involving
illuminated clumps of disk debris ($\S$~2.2), stars ($\S$~2.3), or {\em
any that involves clumpy, localized emission entities}.

\subsection{Accretion Disks \& their Debris}

Given that the quasar paradigm involves a disk of matter accreting onto
a supermassive black hole, with the inner regions emitting the optical
-- X-ray continuum (Rees 1984), a minimalist approach would put the
line emitting gas in the outer regions of the accretion disk. Nature
already does this in stellar cataclysmic variable systems. The
existence of cool, high column density material in the vicinity of the
X-ray continuum source is supported by observations in Seyfert~1 nuclei
of hard X-ray reflection humps and broad iron K$\alpha$ emission lines
near 6.4~keV, whose line profiles obtained by ASCA can be fitted well
by some accretion disk models (Tanaka et al.\ 1995; Fabian et
al.\ 1995; Mushotzky et al.\ 1995; Iwasawa et al.\ 1996; Nandra et
al.\ 1997; Reynolds et al.\ 1997). These reflection components are
rarely observed in quasar spectra, however. A further advantage of a
disk like geometry is that the probability of one or more broad line
emitting entities lying along the line of sight becomes small for
random orientations, and very few quasars or AGN show evidence of
``cold'' X-ray absorption, expected from a typical broad line
``cloud''. Finally, there is no confinement problem for gas associated
with accretion disk, and the gas reservoir is large. The primary
drawback of the accretion disk scenario is that at present its
structure and mechanism for energy dissipation are not well understood.
Most models of accretion disk structure are based upon the work of
Shakura \& Sunyaev (1973). See Blandford (1985), Begelman (1985), and
Frank, King, \& Raine (1992) for good reviews.

\subsubsection{Illuminated Disks.}
Hard X-ray illumination\footnotemark\/ of the outer regions ($10^2 R_g
< r < 10^3 R_g$; here $R_g = 2GM_{bh}/c^2$) of the accretion disk has
been proposed by Collin-Souffrin and co-workers (Collin-Souffrin,
Hameury, \& Joly 1988a; Dumont \& Collin-Souffrin 1990; Rokaki et
al.\ 1993) as a means to account for the low ionization broad emission
lines: Balmer lines, Mg~II~2800, and the forest of Fe~II lines. These
lines are powered via the hard X-ray heating of the dense ($n_H \sim
10^{12-14}$~cm$^{-3}$), high column density ($N_H > 10^{25}$~cm$^{-2}$)
medium. Under these conditions, the emission is not particularly
sensitive to the structure of the lower ``chromosphere'' of the disk
where these lines are emitted.  However, the medium ionization (e.g.,
C~III]~1909) and high ionization (e.g., C~IV~1549) lines, including
Ly$\alpha$, derive their energy from softer ionizing photons (13.6~eV
-- 200~eV), and their emission is very sensitive to the uncertain
structure of the upper disk chromosphere, if they are emitted there at
all.

\footnotetext{The geometry of the innermost accretion disk and the
X-ray emitting continuum source are not understood, thus the mechanism
governing the X-ray illumination of the disk is not well constrained.}

\subsubsection{Illuminated Disk Debris.}

Radiation pressure driven winds from accretion disks have been most
recently investigated by Murray \& Chiang (1997; see also Yamamoto in
these proceedings). Resonance line scattering provides most of the
pressure to accelerate the wind, and the gas densities are kept low to
minimize the mass load of the radiatively powered wind.  Substantial
filtering of the soft X-ray continuum must occur within ``hitchhiking
gas'' just interior to the wind, or else the gas would be overionized
and the wind would die. The bulk of the broad line emission occurs in
regions near the base of the wind close to the disk surface where the
wind radial velocity component is small compared to the disk Keplerian
velocity component, but the radial and rotational velocity shears are
large.  These shears induce anisotropic photon escape probabilities,
and the radial shear component can produce singly peaked emission lines
{\em even though the velocity field is dominated by Keplerian
rotation.} The particle densities of the line emitting gas range from
$10^7$~cm$^{-3}$ to $10^9$~cm$^{-3}$, much lower than all other modern
models of the BELG and their spectra (e.g., Ferland et al.\ 1992). A
possible shortcoming is that such gas is unlikely to produce the
observed Balmer lines, Mg~II~$\lambda$2800, or the Fe~II spectrum. This
model has not yet been computed self consistently (true enough of all
dynamical BLEG models) in that the gas temperature, ionization, and
optical depths are computed as if in a static slab, whereupon the line
escape probabilities are altered for single scattering within the
velocity shears within the wind flow. The geometry of an accretion disk
wind with central and local sources of photons is not well understood,
either.

Another model for disk debris is described most recently in Cassidy \&
Raine (1996). A nuclear wind of unspecified origin interacts with a
thermally driven wind generated by accretion disk gas heated to the
Compton temperature by the central X-ray source. The interface between
these winds near the accretion disk generates Kelvin-Helmholtz
instabilities that create dense clouds that are then accelerated in the
flow before being destroyed. The velocity field of the line emitting
gas is Keplerian flow dominated upon injection and takes on larger
radial contributions as the cloud is accelerated in the wind. High and
low ionization lines are formed in the debris. The hot wind should have
temperatures of order $10^8$~K, and if the quasar radiation field
cannot do this, then some other mechanism must provide the necessary
heating.

A final scenario involving debris from accretion disks has magnetically
confined ``clouds'' centrifugally accelerated above the accretion disk
(Emmering, Blandford, \& Shlosman 1992; K\"{o}nigl \& Kartje 1994;
de~Kool \& Begelman 1995; Bottorff et al.\ 1997). Magnetic lines of
force emanate outwards and upwards from the disk, and blobs of matter
are loaded onto the field lines and then centrifugally accelerated.
Once exposed to the central ionizing source these ``clouds'' are
photoionized and produce emission lines before dissipating at the
Compton temperature at high altitudes above the disk. The velocity
field is cylindrical, with contributions from Keplerian and z
components varying with distance along and above the disk. Given the
thermal, radiation, and magnetic pressures present, the emitting
entities may be shaped like spaghetti or fetuccini (Bottorff \&
Shlosman 1998), and both high and low ionization lines are formed in
the distribution of pasta shaped clouds.

\subsubsection{Summary.} No model involving disks yet stands high above
the others, and all suffer from our ignorance of the detailed structure
of quasar accretion disks.

\subsection{Stars \& their Debris}

Certainly Seyfert~1 nuclei and most and perhaps even all quasars are
associated with the nuclei of galaxies (see Korista, these proceedings,
for references), composed of stars and an ISM.  The center of our
Galaxy is a relatively active region consisting of young star clusters
and very dense molecular clouds.  A reasonable question to ask is what
would happen if one placed an accreting supermassive black hole down in
the center of this?  A number of ideas have suggested a marriage of the
stellar and interstellar environment to the quasar paradigm.

\subsubsection{Bloated Stars.}

A number of investigators have proposed ``bloated stars'' in one form
or another for the origin of the BLEG (Edwards 1980; Penston 1988;
Scoville \& Norman 1988; Kazanas 1989; Baldwin et al.\ 1996), though it
wasn't until the work of Alexander \& Netzer (1994) that the model was
investigated in greater detail. The idea is that a dense star cluster
lies and evolves in the vicinity of the supermassive black hole of the
quasar with its source of ionizing continuum photons (Murphy, Cohn, \&
Durisen 1991). The cross sections presented to the illuminating
continuum source by normal cluster giant or supergiant stars are
insufficient to produce the observed line emission, since the expected
numbers of accompanying dwarf stars would otherwise be too high to
allow for a quasi-stable star cluster. So instead the stars are assumed
to be ``bloated'' through normal stellar winds and/or those induced by
the hard photon flux or even line driving (Baldwin et al.\ 1996). These
winds are likely anisotropic, perhaps somewhat akin to``comet tails''
(e.g., Scoville \& Norman 1995), though this structure has never been
computed fully self consistently. The enhanced cross sections are
sufficient for efficient reprocessing of the continuum into the broad
emission lines.  Stars might also wander within the black hole's tidal
radius and be shredded (Roos 1992), but the tidal radius occurs outside
the Schwarzschild radius only when the mass of the central objects is
less than about $10^8~M_{\odot}$, unless the star's average density is
substantially sub-solar.  The stars themselves are self gravitating and
in any case provide a huge reservoir of gas upon which the BLEG may
draw, and the work of Alexander \& Netzer suggests that only
$\sim$~50,000 bloated stars are required in the broad line emitting
environment of a moderate luminosity quasar.

\subsubsection{Hybrids: Star -- Disk Interaction Models.}

To account for the higher ionization lines that may not be emitted in
their illuminated accretion disk model, Collin-Souffrin et al.\ (1988b)
proposed a hybrid disk/stellar debris model. A nuclear star cluster
produces supernova and stellar wind ejecta that are shocked in a dense
ISM and cool to moderately high densities ($n_H \sim
10^{9-12}$~cm$^{-3}$) (Perry \& Dyson 1985; Williams \& Perry 1994) in
the vicinity of the quasar accretion disk. This fragmenting ``stellar
shrapnel'' is transient and replenished by the stellar cluster, and so
does not require a confinement mechanism. Another hybrid scenario for
the origin of the BLEG involves the stars of a dense nuclear star
cluster colliding with the accretion disk and the resulting ``star
tails'' become the line emitting gas (Zurek, Siemiginowska, \& Colgate
1994).

\subsubsection{Summary.} In principle, the ``bloated'' stars scenario
should presently be on firmer theoretical footing than other scenarios
for the BLEG (e.g., accretion disks), since the structure of normal
stars is well understood.  Even so, no self-consistent models for a
stellar atmosphere's interaction with the full quasar environment have
been computed.  The primary concerns of the stellar origin scenario of
the BLEG have been generally related to the question of packing enough
stars close enough to the quasar's ``central engine.''

\section{Evidence}

By now the reader should be convinced that the broad range in BLEG
origin scenarios is indicative of a relative ``vacuum'' in the
observational constraints. In addition, gaps in our understanding of
some of the physics allow for a plethora of free parameters. While all
true, progress has certainly been made on both fronts, especially since
the advent of the solid-state detector spectrograph some 20 years ago,
and with the rapid increase in computational power in the same time
period.  The following is a non-exhaustive review of the observational
evidence that constrains the origin and nature of the BLEG.

\subsection{Continuum -- Line Reverberation}

Since the emission lines are powered primarily or exclusively via
photoionization, a time-variable ionizing continuum flux will be seen
by a distant observer to reverberate throughout the BLR as
time-variable emission lines. In principle the structure and dynamics
of the emitting gas can then be derived (Blandford \& McKee 1982;
Capriotti, Foltz, \& Peterson 1982; Welsh \& Horne 1991; P\'{e}rez,
Robinson, \& de la Fuente 1992). The recovered structure and dynamics
might then point to an origin of the BLEG or exclude others.  However,
recovery of this information via reverberation has proven difficult.
Combinations of uneven, insufficiently frequent, and insufficiently
extensive time sampling have hampered our efforts. An extensive
literature exists on this subject and will not be reviewed here. I
refer the reader to Peterson (1993; 1997) for recent reviews of the
subject. A summary of the information gained concerning the BLR through
these experiments is as follows. (1) The BLR is geometrically extended,
and its characteristic distance from the ionizing continuum source
scales as $\sim 0.1 L_{46}^{1/2}$~pc, where $L_{46}$ is the ionizing
continuum luminosity in units of $10^{46}$ ergs/s. (2) The BLR is
stratified in ionization, and a range of gas densities must be
present.  (3) Pure radial flows, inward or outward do not dominate the
BLEG dynamics.  Any model for the origin of the BLEG must fall within
these constraints.

\subsection{Emission Line Profiles}

If the emission line profiles were to suggest a particular gas
dynamics, this in turn might constrain the origin of the BLEG
(Blumenthal \& Mathews 1975; Capriotti, Foltz, \& Byard 1981a).
Unfortunately, the emission line profiles, characterized roughly as the
functions $I_{\lambda} \propto \log(\Delta\/\lambda\/)$ and/or
$I_{\lambda} \propto (\Delta\/\lambda\/)^{-2}$, do not point to a
unique dynamical mechanism for the emitting gas, though their forms are
indeed constraints on any model. Most emission lines are singly peaked,
though are often accompanied by bumps or shoulders, and are often
asymmetric (e.g., see Boroson \& Green 1992). Also unexplained are the
observed emission line peak redshift differences, sometimes exceeding
$\Delta\/v \approx 1000$ km/s (Gaskell 1982; Wilkes 1984; Wilkes 1986;
Espey et al.\ 1989; Corbin 1990; Tytler \& Fan 1992). The higher
ionization lines tend to lie at smaller redshifts, and are likely
blueshifted relative to the lower ionization lines that lie closer to
the quasar's systemic redshift.

Line emission from accretion disks is expected to result in
double-peaked emission lines (see Marsh \& Horne 1986 for a nice
review), and the paucity of such examples in Seyfert~1 and quasar
spectra has been used as an argument against an accretion disk origin
of the broad emission lines (e.g., see Sulentic 1989; Sulentic et
al.\ 1990). However, the anisotropic line radiative transfer expected
in radiatively driven winds (Murray \& Chiang 1997) and the 2-D
velocity field nature of the magnetically driven ``winds'' (Emmering et
al.\ 1992; Bottorff et al.\ 1997) can produce singly peaked emission
lines. In any case a small number of objects (e.g., Arp~102B, 3C~390.3,
3C~332, and others) do exhibit broad, double-peaked emission lines
(Chen \& Halpern 1989; Eracleous \& Halpern 1994; Halpern et
al.\ 1996). These objects tend to be LINERs (``low ionization nuclear
emission regions''; e.g., see Barth et al.\ 1998) and/or radio
galaxies, usually with weak or absent UV
bumps\footnotemark\/.\footnotetext{I find it ironic that the objects
that exhibit the classical double-peaked accretion disk emission line
profiles lack the classical UV continuum signature of accretion disks
(Shields 1978; Malkan \& Sargent 1982).} Generally, only the Mg~II
$\lambda$2800 and Balmer emission lines exhibit the double-peaked
profiles. Halpern et al.\ (1996) argued that the lack of a double-peak
signature in Ly$\alpha$~$\lambda$1216 in some objects indicates
thermalized emission such as that expected from high density and column
density gas in accretion disks. In an effort to confirm or disprove the
accretion disk hypothesis for the BLEG origin of this subclass of
objects, a number of studies have sought to carefully measure the
emission line profiles and monitor their variations due to
reverberation effects. Some of the observed profiles and their
variations were unexpected based upon simple models of line emitting
accretion disks (Miller \& Peterson 1990; Veilleux \& Zheng 1991;
Zheng, Veilleux, \& Grandi 1991; Sulentic et al.\ 1995; Eracleous et
al.\ 1997). However, effects such as disk ``hot spots'' (Zheng,
Veilleux, \& Grandi 1991; Newman et al.\ 1997) or two-armed spiral
waves\footnotemark\/ (Chakrabarti \& Wiita 1994)\footnotetext{Steeghs,
Harlaftis, \& Horne (1997) and others have identified signatures of
line emission from enhanced gas density via spiral shocks in the disks
of cataclysmic variable systems.}, or precessing elliptical disks
(Storchi-Bergman et al.\ 1997) might explain some of the anomalies.

Three historically narrow-line emitting objects (NGC~1097, M~81, and
Pictor~A) recently and suddenly exhibited broad, double-peaked Balmer
emission lines.  It has been suggested that this could arise from the
sudden appearance of the broad line region (clearing of obstruction),
the tidal disruption of at least one star, or the sudden switch-on of a
source of high energy photons in the presence of an established
accretion disk (Storchi-Bergman et al.\ 1995). Their relationship to
the above class of double-peak emitters is unclear.

In several objects with either double peaked or double shouldered
Balmer lines, Zheng, Binette, \& Sulentic (1990), Marziani, Calvani, \&
Sulentic (1992), and Sulentic et al.\ (1995) find that the profiles and
their variations are better fitted by two-stream or bi-cone
subrelativistic flow models. The nature of these flows and thus of the
emitting gas, however, has been left generally unspecified by these
investigators.

\subsection{Other Considerations}

An often tacit assumption in simulations of quasar emission line
spectra (e.g., Rees, Netzer, \& Ferland 1989) is that the local line
width is dominated by thermal gas motions. This may occur in confined
or quasi-static clouds, but in most other scenarios for the BLEG the
emitting gas is locally dynamic, and significant local line broadening
can result. This alters the diffuse emission escape probability as
well as the ionization and thermal structures of the emitting gas, and
a significant contribution from continuum pumping becomes possible
(Shields, Ferland, \& Peterson 1995). In the near future, as computers
become fast enough to simulate accurately the radiative transfer of the
diffuse emission, we should consider testing the effects of
extra-thermal local line broadening on the computed emission line
spectrum. We may be able to set some limits to the presence of velocity
gradients, shears, or streams. These theoretical limits may then be
tested against the emission line profile smoothness requirements from
observations (Arav et al.\ 1998a,b), and perhaps then we can say what
the BLEG is not.

One of the most interesting and promising clues from quasar emission
line spectra is that they represents emission from chemically enriched
gas with metallicities broadly constrained to lie within
1--10~$Z_{\odot}$, even in the highest redshift sources (Hamann \&
Ferland 1993; Hamann \& Ferland 1999; Hamann, these proceedings). Rapid
star formation and evolution in dense cluster environments can account
for these enrichments, although other scenarios are possible. For
further discussion of quasar gas metallicity, see my other conference
proceedings contribution. Any successful model for the origin of the
BLEG will need to address how the enriched gas became the broad line
emitting gas.

\section{Summary \& Comments}

{\em Anyone who isn't confused just doesn't understand what's going
on.

--- anonymous}\\

\noindent
One of the outstanding mysteries concerning quasars and active galaxies
is the origin and nature of the broad line emitting gas. No model or
scenario fits all the observational data, and most suffer from too few
observational constraints and too many free and often {\em
ill-understood} parameters.

Unless we figure out how to image the central regions of active
galaxies, the continuum and emission line reverberation experiments
offer the most powerful means to placing useful constraints upon the
nature of the BLEG and possibly its origin.  However, it will likely
take a dedicated multiwavelength spectroscopic orbiting platform before
we can learn anything further from reverberation.

A large majority of what we do know about quasars arises from their
spectra. However, some of our observational constraints are muddied by
spectra of moderate or even poor quality. {\bf The establishment of a
central data base of high quality (S/N, spectra resolution and
coverage) quasar spectra} and their uniform measurements would lay a
firm foundation of observational constraints upon which we can build
our understanding of quasar spectra. One century ago, such a data base
existed for stars (e.g., Harvard Observatory), where they were
classified ``zoologically'' by characteristics of their spectra.
However, the nature of stars and their spectra were unclear. With the
dawning of the $20^{th}$ century came our realization of the quantum
nature of light (Planck and Einstein), the theoretical progress set
forth by Saha and Eddington and others, and the identification of the
stellar main sequence by Hertzsprung and Russell. The second decade
gave birth to quantum mechanics and the salient Ph.D.\ thesis of
Cecilia Payne, and finally we understood the story of stellar spectra.
The third through sixth decades saw the development of the theory of
nucleosynthesis by Bethe, Gamow, then Fowler, Hoyle, and Geoffrey and
Margaret Burbidge (the latter two also important investigators of
quasars), and then Iben's work on stellar evolution. {\em Underlying
all of this progress was a vast stellar spectroscopic data base.} One
of the greatest accomplishments of this century has been the
understanding of stars. Today we stand as we did one century ago,
perhaps too, near a precipice. Will the next century bring a similar
leap in our understanding of quasars and their place in the universe?

\acknowledgments
I thank Jack Baldwin and Gary Ferland for giving me the opportunity to
write this review.

\end{document}